\documentclass[runningheads]{llncs}

\usepackage[noend]{algpseudocode}
\usepackage{algorithm}

\usepackage{graphicx}
\usepackage{subfigure}
\usepackage{amsmath,amssymb,amsfonts}
\usepackage{color}
\usepackage{ulem}
\usepackage{hyperref}
\hypersetup{
    colorlinks=true,
    linkcolor=blue,
    filecolor=magenta,      
    urlcolor=cyan,
    pdftitle={Overleaf Example},
    pdfpagemode=FullScreen,
    }
\DeclareMathOperator*{\argmin}{arg\,min}

\newcommand{\norm}[1]{\left\lVert#1\right\rVert} 

\makeatletter
\newcommand{\printfnsymbol}[1]{%
  \textsuperscript{\@fnsymbol{#1}}%
}
\makeatother

\begin{document}
\title{Interpretable Modeling and Reduction of Unknown Errors in Mechanistic Operators} 

\titlerunning{Interpretable Modeling and Reduction of Unknown Errors in Mechanistic Operators}

\author{Maryam Toloubidokhti $^*$ \inst{1} \and Nilesh Kumar $^*$ \inst{1} \and Zhiyuan Li \inst{1} \and
Prashnna K. Gyawali\inst{2} \and Brian Zenger\inst{3} \and Wilson W. Good\inst{3}
\and Rob S. MacLeod \inst{3}
\and Linwei Wang\inst{1} 
} 
%
%
\institute{Rochester Institute of Technology, NY, USA\\
\email{mt6129@rit.edu} \\
\email{nk4856@rit.edu}
\and{Stanford University}
\and{The University of Utah, UT, USA}}

\maketitle             

\begin{abstract}

Prior knowledge about the imaging physics 
provides a mechanistic \textit{forward} operator that 
plays an important role in image reconstruction, although myriad sources of possible errors in the operator could negatively impact the reconstruction solutions. 
In this work, 
we propose to embed the  traditional  mechanistic  forward  operator inside a neural function, and focus on 
modeling and correcting its unknown errors
in an interpretable manner. 
This is achieved by a conditional generative model that transforms a given mechanistic operator 
with unknown errors, 
arising from a latent space of self-organizing clusters 
of potential sources of error generation. 
Once learned, 
the generative model can be used in place of a fixed forward operator in any traditional optimization-based reconstruction process where, 
together with the inverse solution, 
the error in prior mechanistic forward operator can be minimized and the potential source of error uncovered. 
We apply the presented method to 
the reconstruction of heart electrical potential from body surface potential. 
 In controlled simulation experiments and \textit{in-vivo} real data experiments, we demonstrate that the presented method allowed reduction of errors in the physics-based forward operator and thereby delivered inverse reconstruction of heart-surface potential with increased accuracy.
\end{abstract}

\section{Introduction}

\def\thefootnote{*}\footnotetext{These authors contributed equally to this work}

In traditional approaches to medical image reconstruction, prior knowledge about the underlying imaging physics plays an important role. We typically start with a \textit{mechanistic forward operator} that defines the relationship between the unknown inverse solution and the external measurements, obtained by numerically solving mathematical equations governing the underlying imaging physics.
The inverse problem of image reconstruction is then formulated as deterministic optimization or probabilistic inference leveraging these mechanistic forward operators \cite{formaggia_quarteroni_veneziani_2006,natterer2001mathematical}. Errors in these operators, arising from factors such as assumptions and imperfect knowledge associated with the underlying physics, heavily affect the optimization or inference processes
\cite{formaggia_quarteroni_veneziani_2006}. 
Due to myriad sources of possible errors associated with the mechanistic operator, it is also non-trivial to attempt to directly minimize or correct these errors during the inverse process. The sources of error we considered represent the most prominent sources of errors in ECG imaging as concluded in literature \cite{476126}. These errors arise from different stages in the ECG imaging pipeline such as data acquisition, image segmentation, and forward modeling.

Alternatively, modern deep learning has shown great potential in medical image reconstruction across a variety of modalities \cite{8253590,ramanarayanan_murugesan_ram_sivaprakasam_2020,DeepPET,Ghimire}. 
These approaches do not rely on the correctness of any prior physics knowledge and directly learn the forward and/or inverse mapping from data. The neglect of available knowledge however must be compensated by a large number of labeled training pairs, which are often difficult to obtain in image reconstruction problems. Furthermore, 
the data-driven black-box reconstruction process leads to limited interpretability.

There has been substantial interest in combining the strength of deep learning with available prior knowledge of the imaging physics. For instance, in a traditional formulation of reconstruction problems utilizing mechanistic forward operators, existing works have explored the possibility to learn better regularizers for the optimization \cite{MRI_ODE}, or to learn the optimization gradient itself \cite{Proximal}. 

In this work, we are motivated to 
retain our prior knowledge of the known imaging physics, while 
focusing on 
modeling and correcting its unknown errors in an interpretable manner. 
To this end, 
we propose a hybrid mechanistic and neural modeling of the forward operator: given a mechanistic forward operator $\mathbf{H}_{i}$ obtained for instance by solving partial differential equations governing the imaging physics, 
we introduce 
 a conditional generative model $\mathbf{H}_{f} = \mathbf{G}(\mathbf{H}_{i},\mathbf{z}_r)$ that,
 in the form of a U-net \cite{UNET}, 
 combining $\mathbf{H}_{i}$ and any unknown residual errors generated from the latent variable $\mathbf{z}_r$ as illustrated in Fig.~\ref{fig:overview}.

Furthermore, 
we describe $\mathbf{z}_r$ with self-organizing maps (SOM) \cite{kohonen1990self} to provide an interpretable model of 
the various potential sources that generate the errors.
This hybrid and interpretable generative model is optimized via 
amortized variational inference. 
Once learned, 
the generative $\mathbf{G(H}_{i},\mathbf{z}_r)$ can be used in place of a fixed forward operator in any traditional optimization-based reconstruction process where, 
together with the inverse solution, 
the error in prior mechanistic forward operator can be minimized and the potential source of error uncovered 
(\textit{e.g.}, to shed light on future data collection for a more accurate forward modeling).

We demonstrate this novel concept on the reconstruction of cardiac electrical potential from body surface potential \cite{Gulrajani,Horek1997TheIP}. 
A similar motivation was pursued in a previous work in this specific application (DAECGI) \cite{MTBDAECGI2021}, 
where a conditional generative model was developed to learn the forward operator 
as a function of known geometry underlying the operator  
and an unknown latent variable. 
As a result, the generative model is not hybrid with the mechanistic forward operator, but mimics the mechanistic operator with a fully data-driven neural function of the input geometry. This modeling option also limits it to the particular application where the forward operator is a function of the underlying geometry. Furthermore, there was not an attempt to provide an interpretable modeling of the sources of error generation.
In controlled simulation experiments and \textit{in-vivo} real data experiments, we demonstrate that the presented method allowed
reduction of errors in the mechanistic forward operator and 
recognition of the sources of errors. Furthermore, we demonstrated that the presented method improved the accuracy of the reconstructed heart-surface potential, in comparison to DAECGI as described in \cite{MTBDAECGI2021}.

\section{Method}
\begin{figure}[!tb]
    \begin{center}
        \includegraphics[width=0.8\textwidth,scale=0.4]{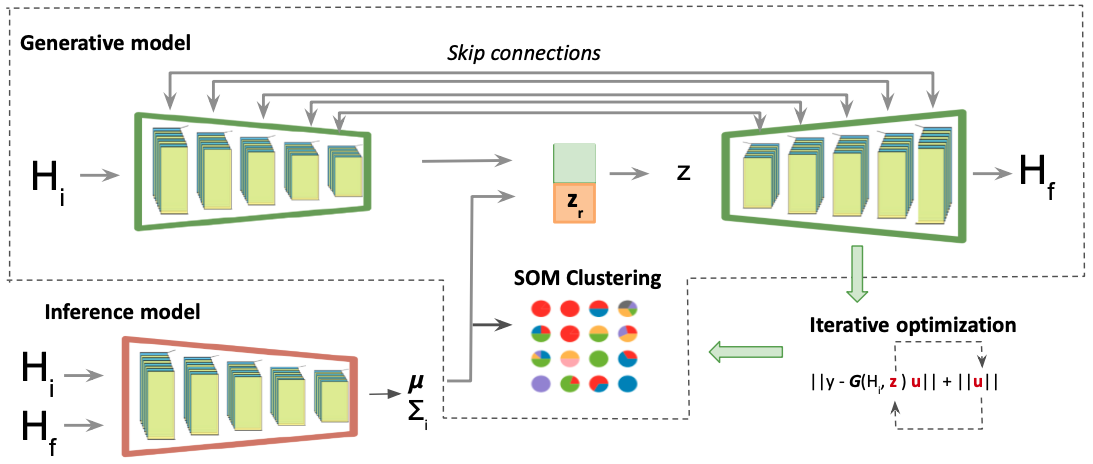}
        \caption{Overview of IMRE network. Top: generative model; Bottom left: inference model; Bottom center: SOM clustering; Bottom right: Iterative optimization and reduction of errors in the generative model. }
        \label{fig:overview}
    \end{center}
\end{figure}

As an application context of the proposed methodology, 
we focus on the reconstruction of cardiac electrical
potential from body surface potential, 
also known as electrocardiographic imaging (ECGi). 
Considering the extracellular potential $\mathbf{u}$ on the heart surface as the electrical source, its relationship with body surface potential $\mathbf{y}$ is defined by the Laplace's equation \cite{plonsey_fleming_1989}. Numerically solving this Laplace equation 
on the heart-torso geometry of any subject
gives rise to the forward mechanistic operator $\mathbf{H}$ that relates heart surface potential to body-surface potential.
The reconstruction of the $\mathbf{u}$ can then be formulated as: 

\begin{equation}
\label{eq:optim_eqn_1}
    \hat{\mathbf{u}} = \argmin_\mathbf{u} \{||\mathbf{y} - \mathbf{H}\mathbf{u}|| + \lambda\mathcal{R}(\mathbf{u})\}
\end{equation}
 where  $\mathcal{R}(\mathbf{u)}$ is the regularization term weighted by parameter 
 $\lambda$.

We present Interpretable Modeling and Reduction of Unknown Errors in Mechanistic Operators (IMRE) for solving Eq.~(\ref{eq:optim_eqn_1}) which, as outlined in Fig.~\ref{fig:overview}, consists of two components: 1) a generative model, shown in the dashed box, that models the unknown errors in the mechanistic forward operator in an interpretable manner, 
and 2) an iterative optimization scheme to jointly optimize the
inverse solution and minimizes the error in the
forward operator. 
Once optimized, 
we can further identify the potential source of errors 
owing to the use of SOM in the latent space of error modeling. 
 
\subsection{Interpretable Modeling of Unknown Errors in Prior Physics}
The presented generative model respects the mechanistic forward operator
and focuses on learning its potential errors, 
such as those caused by inaccuracy in the underlying heart-torso geometry or tissue conductivities. It is further enriched with a interpretable latent-space modeling to cluster possible sources of errors.

\subsubsection{Generative model:}

As outlined in the dashed box in Fig.~\ref{fig:overview}, 
generation of the hybrid forward operator is 1) conditioned on the 
mechanistic forward operator $\mathbf{H}_{i}$ via a U-NET structure, and 2) augmented with a latent representation $\mathbf{z}_{r}$ that explains residual errors in the prior physics: 
\begin{equation}
\label{gen_model}
\mathbf{H}_f \sim 
p(\mathbf{H}_{f}|\mathbf{H}_{i}, \mathbf{z}_{r}), \quad
p(\mathbf{z}_r) \sim 
\mathcal{N}(\mathbf{0},\mathbf{I})
\end{equation} 

The U-NET provides direct links from $\mathbf{H}_{i}$ to the decoder, allowing us to preserve the prior physics and focus on modeling its potential errors.

\subsubsection{Self Organizing Maps (SOM):}
Considering that these errors can arise from several different sources, 
we further group sources of error $\mathbf{z}_{r}$ with a SOM \cite{kohonen1990self}. SOM is an unsupervised machine learning algorithm that makes certain parts of the network respond similarly to the matching input patterns. We consider a SOM that consists of a set of $\mathbf{K}$ nodes $\mathbf{\mathcal{V}=\{v_1,v_2,...,v_{K}\}}$ 
on a two-dimensional gird. To model $\mathbf{z}_r$, during training, a weight vector $\mathbf{w}_v$ is learned for each node $\mathbf{v}$ to store a specific type of the latent variable $\mathbf{z}_{r}$. Given the distribution of $\mathbf{z}_r$, its best matching unit (BMU) on SOM is found by:
\begin{equation}
  v_{\text{BMU}}=\mathrm{argmin}_{v\in \mathcal{V}}\norm{\mathbb{E}[\mathbf{z}_r]-\mathbf{w}_{v}}^2,
\end{equation}
Based on the BMU, weight vectors of each node $v\in \mathcal{V}$ (including the BMU) can be updated as:
\begin{equation}
  \mathbf{w}_{v}\gets \mathbf{w}_{v}+\gamma N(v,v_{\text{BMU}})(\mathbb{E}[\mathbf{z}_r]-\mathbf{w}_{v_{\text{BMU}}}),
\end{equation}
where $\mathbf{\gamma}$ is the learning rate, and $\mathbf{N(v,v_\text{BMU})}$ is a neighborhood function (\textit{e.g.}, Gaussian or triangular) that has higher value when $\mathbf{v}$ and $\mathbf{v_\text{BMU}}$ are closer on SOM, which can be loosely determined by SOM rules of thumb.
At test time, the embedding of the test samples in the latent space are used to find the BMU in the SOM and assigning the error source label as discussed in Section ~\ref{SOM-sec}. 

\subsubsection{Variatioanl Inference:}

We derive the modified evidence lower bound (ELBO) of the log conditional likelihood of $p(\mathbf{H}_{f}|\mathbf{H}_{i})$ with $\beta$ parameter as: 
\begin{equation}
\label{eq:var_infr}
\begin{split}{}
\log p(\mathbf{H}_{f}|\mathbf{H}_{i}) \ge \mathcal{L}_{ELBO} = 
    E_{\mathbf{z}_r\sim q(\mathbf{z}_r|\mathbf{H}_{i}, \mathbf{H}_{f})} p(\mathbf{H}_{f}|\mathbf{z}_{r},\mathbf{H}_{i}) \\
    -  \beta {D}_{KL}(q(\mathbf{z}_{r}|\mathbf{H}_{i}, \mathbf{H}_{f})||p(\mathbf{z}_{r} ))
\end{split}    
\end{equation} 
where we add a hyperparameter $\beta$ to the second term. 
We use a variational distribution $q(\mathbf{z}_{r}|\mathbf{H}_{i}, \mathbf{H}_{f})$, parameterized by an inference network as illustrated in the brown encoder in Fig.~\ref{fig:overview}, to approximate the intractable posterior distribution. 
Mimicking the structure of a conditional variational auto-encoder (VAE), 
the conditioning of the encoder on $\mathbf{H}_{i}$ further encourages $\mathbf{z}_{r}$ to
focus on learning the residual between $\mathbf{H}_{i}$ and $\mathbf{H}_{f}$. Overall, 
The first term in Equation (\ref{eq:var_infr}) encourages the reconstruction of the desired $\mathbf{H}_{f}$ given the mechanistic forward operator $\mathbf{H}_{i}$.
The second term regularizes $q(\mathbf{z}_{r}|\mathbf{H}_{i}, \mathbf{H}_{f})$ with a prior distribution $p(\mathbf{z})$, which we assume to be an isotropic Gaussian.

We also note that, without additional regularization, learning with the above loss can collapse to a trivial solution where the model may ignore the conditioning on $\mathbf{H}_{i}$ and
learn to directly model 
$\mathbf{H}_{f}$ instead. To avoid this, we ask the model to generate $\mathbf{H}_{i}$ if the variational encoder is presented with a pair of identical operators (\textit{i.e.}, no residual to be uncovered). 
This gives rise to a combined loss:

\begin{equation}
\label{eq:loss_fn}
   L = L_{ELBO} + \lambda_{reg}*L_{recon}(\mathbf{H}_{i},\hat{
     \mathbf{H}_{i}}) 
\end{equation}
where $L_{recon}(\mathbf{H}_{i},\hat{
     \mathbf{H}_{i}})$ is the added regularization term with hyperparameter $\lambda_{reg}$.

\subsection{Inverse Estimation with Simultaneous Error Reduction}
When seeking the inverse solution for a given observation $\mathbf{y}$, 
we are interested in 
utilizing the (partially) known 
physics behind the mechanistic forward operator $\mathbf{H}_{i}$ without being bounded by its potential errors.
To do so,  while estimating the inverse solution $\mathbf{u}$, we simultaneously optimize the latent variable $\mathbf{z}_{r}$ in the conditional generative model $\mathbf{G}(\mathbf{H}_{i}, \mathbf{z}_{r})$ to account for the errors in $\mathbf{H}_{i}$. After embedding the conditional generative model in Eq.~(\ref{eq:optim_eqn_1}), it is reformulated as:

\begin{equation}
\label{eq:optim_eqn}
    \hat{\mathbf{u}} = argmin_\mathbf{u}||\mathbf{y} - {\mathbf{G}(\mathbf{H}_{i}, \textbf{z}_{r})}\mathbf{u}|| + \lambda \mathcal{R}(\mathbf{u});
\end{equation}
where $\mathbf{G}(\mathbf{H}_{i}, \mathbf{z}_{r})$ is the mean of $p(\mathbf{H}_{i},\mathbf{z}_{r})$.
As a choice of $\mathcal{R}(\mathbf{u})$ 
in this initial study, we consider the second-order Laplacian operator over the heart surface to enforce the spatial smoothness of the inverse solution. 

In each iteration, the algorithm updates $\mathbf{H}_{f}$ by optimizing $\mathbf{z}_{r}$ with the BOBYQA algorithm \cite{cartis2018improving}, a derivative-free iterative algorithm for finding the minimum of a function within given bounds. This is expected to continuously reduce the error in the initial mechanistic forward operator of $\mathbf{H}_{i}$. With each updated $\mathbf{H}_{f}$, the estimate of $\mathbf{u}$ 
is updated by the second-order Tikhonov regularization. The iterations continue until minimal changes are produced in $\mathbf{u}$ and $\mathbf{H}_{f}$ between consecutive updates, or until the maximum number of iterations is completed.



\subsubsection{Error source detection with SOM:}
\label{SOM-sec}

Once we get the optimized latent values for the possible error in $\mathbf{H}_{i}$, we predict the error type using the trained SOM. Specifically,
given the optimized value of $\mathbf{z}_r$, 
we find its BMU on the SOM map.

We then identify the source of error associated with  $\mathbf{z}_r$ 
using the source label from the majority members associated with its BMU.
In case the node selected was not associated with any training samples, 
we look for the next BMU.   
 
\section{Experiments}
IMRE is trained on mechanistic forward operators generated with controlled errors due to a variety of factors such as translation, rotations, inhomogenity, and scaling of the heart and torso geometries. It is then evaluated using simulated and \textit{in-vivo} real data experiments. IMRE's source code is available at this  \href{https://github.com/miccai2022IMRE/miccai_2022_IMRE}{\textit{Link}.}

\subsection{Datasets}

\subsubsection{Synthetic Data:} 
    We used the open-source SCI-Run software \cite{SCI:SCIRun} to generate 600 mechanistic forward operators using the boundary element method on given heart-torso geometries obtained by transforming a base human heart with rotations along the X-, Y- and Z-axes in the range of [-50°,20°],[-50°,20°], and [-80°,10°], respectively, and translation along X-, Y- and Z-axes in [-60, 60] millimeters. We also performed torso scaling in range of [0.9,1.4] and varying conductivity values in [0,0.13] for introducing inhomogenuity. The forward operators are exhaustively paired and are split into 80-20 ratio for training and testing. To evaluate IMRE in ECGi estimation, we used Aliev-Panfilov (AP) model \cite{ALIEV1996293} to  simulate the extracellular potential considering three different origins of activation and generated the body-surface potential data using $\mathbf{H}_{f}$ with 35DB Gaussian noise. In total IMRE is evaluated for inverse estimation 
    across 75 sets of simulated data.
    
    \subsubsection{Real Data}: 
    We evaluated the trained IMRE on human data samples of body-surface recordings of two CRT patients \cite{potyagaylo2019ecg,Aras2015}, given the ground-truth of the pacing location. We compared the performance of IMRE with DAECGI \cite{MTBDAECGI2021}, and the initial forward operators $\mathbf{H_{i}}$ generated with SCIRun software.

\subsection{Generative Model Training}
We trained the two encoders and the decoder simultaneously for 100 epochs with a batch size of 64 and latent dimension of 16. The value of $\beta$ is set to. 0.001 and $\lambda$ to 0.02. Both encoders have the same  architecture; five convolutional layers followed by four fully connected layers. The decoder consists of three fully connected layers followed by transpose convolutions for up-sampling. We should note that the VAE encoder and decoder are connected through skip-connections as visualized in Fig.~\ref{fig:overview}.    
 
\subsection{Evaluation of Generative Model} 
\begin{figure}[!tb]
    \centering
    \includegraphics[width=0.8\textwidth,scale=0.5]{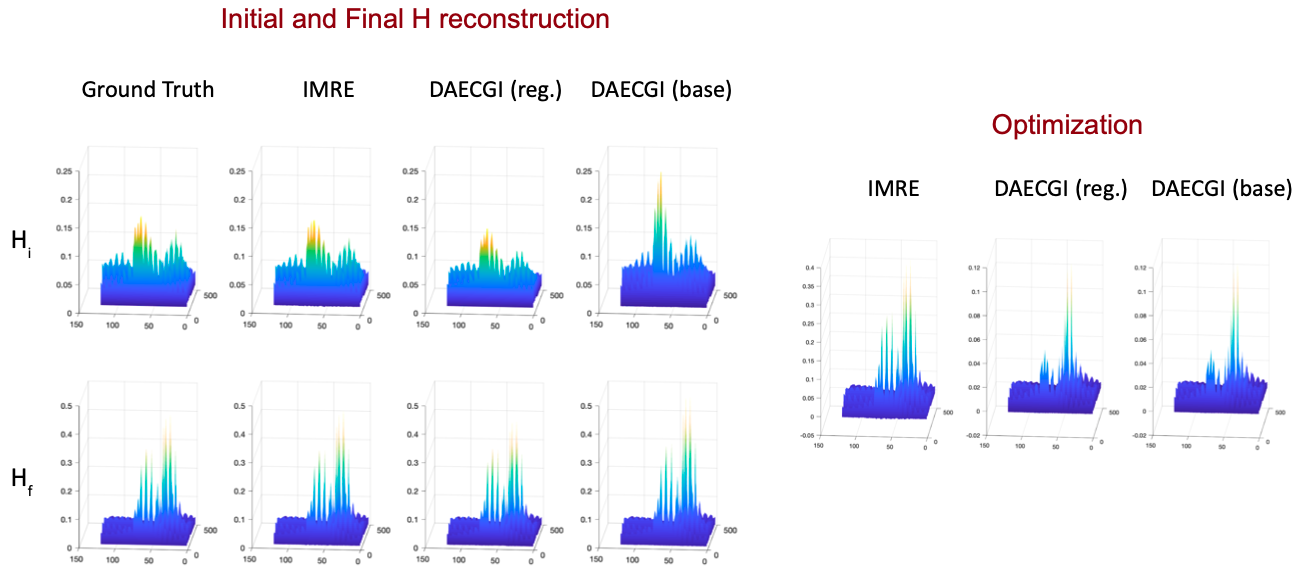}
    \caption{Comparison of generative model accuracy. \cite{MTBDAECGI2021} is referred as DAECGI. Top left row shows reconstruction of $\mathbf{H}_{i}$ given no error, bottom right is $\mathbf{H}_{f}$ reconstruction and center right is optimized forward operator starting from $\mathbf{H}_{i}$.}
    \label{fig:recons}
\end{figure}

IMRE generative model is validated considering the following criteria: 1) successful reconstruction of $\mathbf{H}_{f}$, 2) generating diverse samples for any $\mathbf{H}_{i}$ and different residual errors $\mathbf{z_r}$, and 3) reconstruction of $\mathbf{H}_{i}$ in the absence of error. Fig.~\ref{fig:recons} compares our generative model with DAECGI which fails to reconstruct the $\mathbf{H}_{i}$ if not supplemented with proper regularization. This underscores our point that utilizing both the initial operator $\mathbf{H}_{i}$ and residual errors $\mathbf{z_{r}}$ leads to a more accurate reconstruction of the final forward operator. The RMSE of $\mathbf{H}_{i}$ reconstruction given no error in each model is as follows; IMRE : 0.011, DAECGI with regularization (DAECGI(reg.)): 0.013, DAECGI without regularization (DAECGI(base)): 0.04 and for $\mathbf{H}_{f}$ reconstruction is; IMRE: 0.013, DAECGI(reg.): 0.013, DAECGI(base): 0.014 and the results after optimization is; IMRE: 0.011, DAECGI(reg.): 0.026, DAECGI(base): 0.026.

\subsection{Evaluation of SOM}
\begin{figure}[!tb]
    \centering
    \includegraphics[width=0.5\textwidth,scale=0.3]{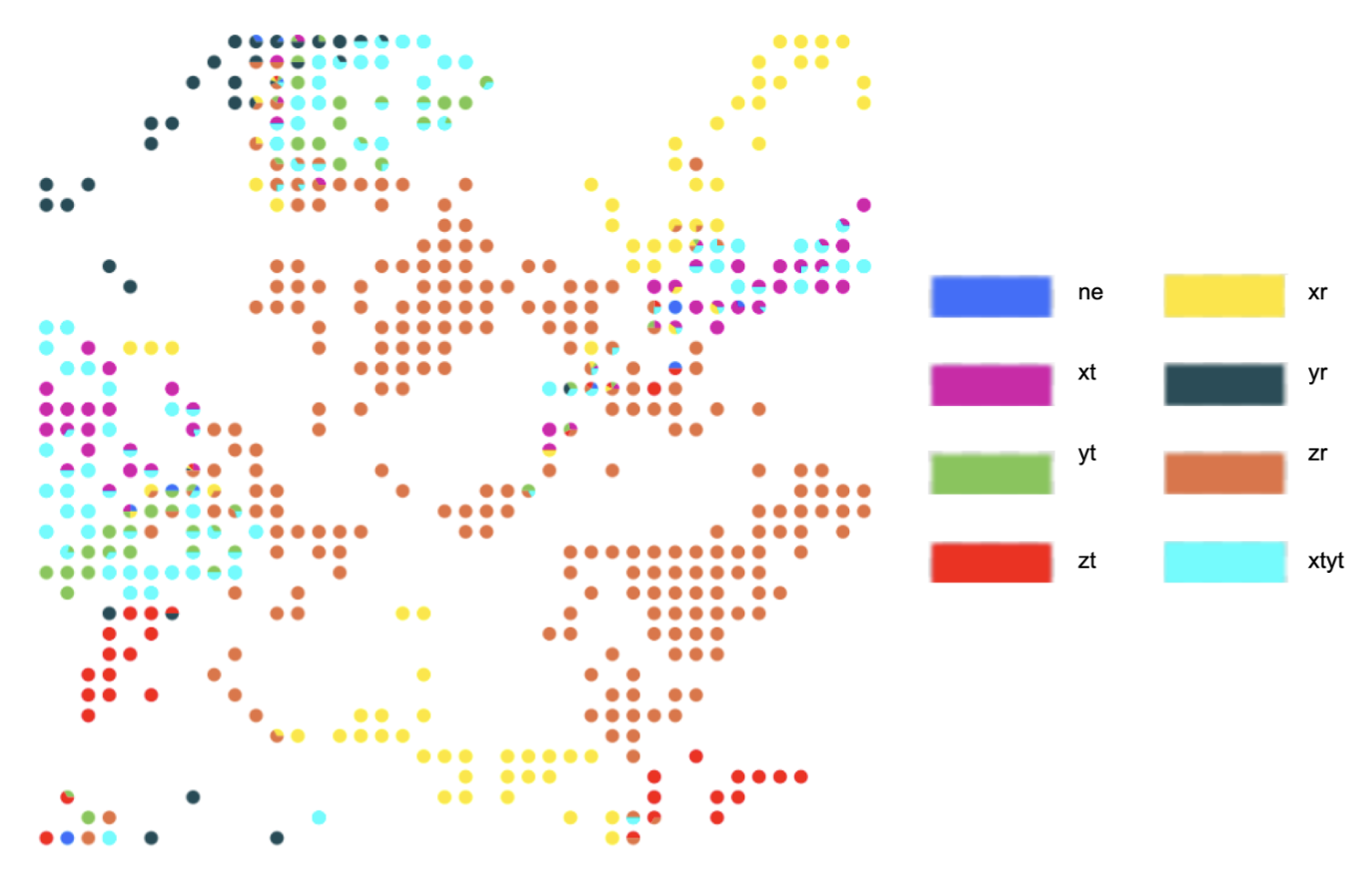}
    \caption{SOM pie chart: color of each node shows the dominant class in that particular node. Number of error classes are reduced for better visibility.}
    \label{fig:som_pie}
\end{figure}

We use the Minisom \cite{vettigliminisom} library for SOM implementation. After tuning the hyper-parameters, we used a SOM grid size of 40x40. We validated the SOM clustering ability by visualizing the clustered data samples in the SOM pie-chart. The clustering performance of this SOM map is crucial to identify the type of errors at test time. It is visible from fig. \ref{fig:som_pie} that the clusters are well-formed. Also, we note that the errors with an overlap in the error type are close to each other in nature. For example, the samples with both error type of x-translation and y-translation sometimes activate the nodes with x-translation or y-translation alone. The classification accuracy on the validation pairs in our simulation dataset is 71\%.

\subsection{Optimization: Inverse Estimation}
\begin{figure}[t!]
    \centering
    \includegraphics[width=0.8\textwidth]{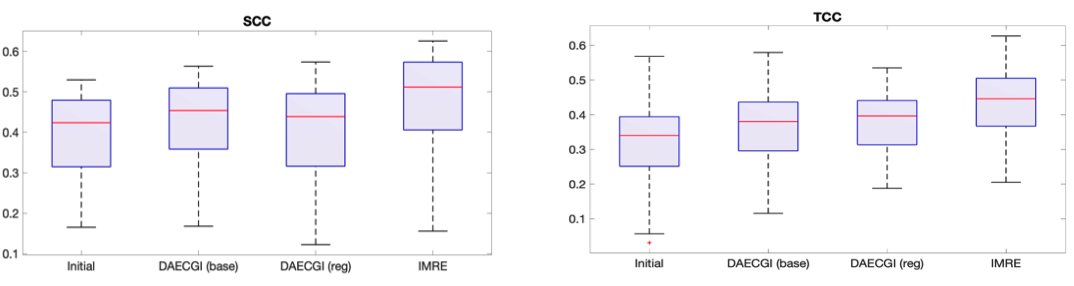}
    \caption{Comparison of inverse reconstruction accuracy. From left to right: initial forward operator, DAECGI, DAECGI ith regularization, and IMRE (Ours).}
    \label{fig:tcc_scc}
\end{figure}

\begin{figure}[t!]
    \centering
    \includegraphics[width=0.5\textwidth,scale=0.2]{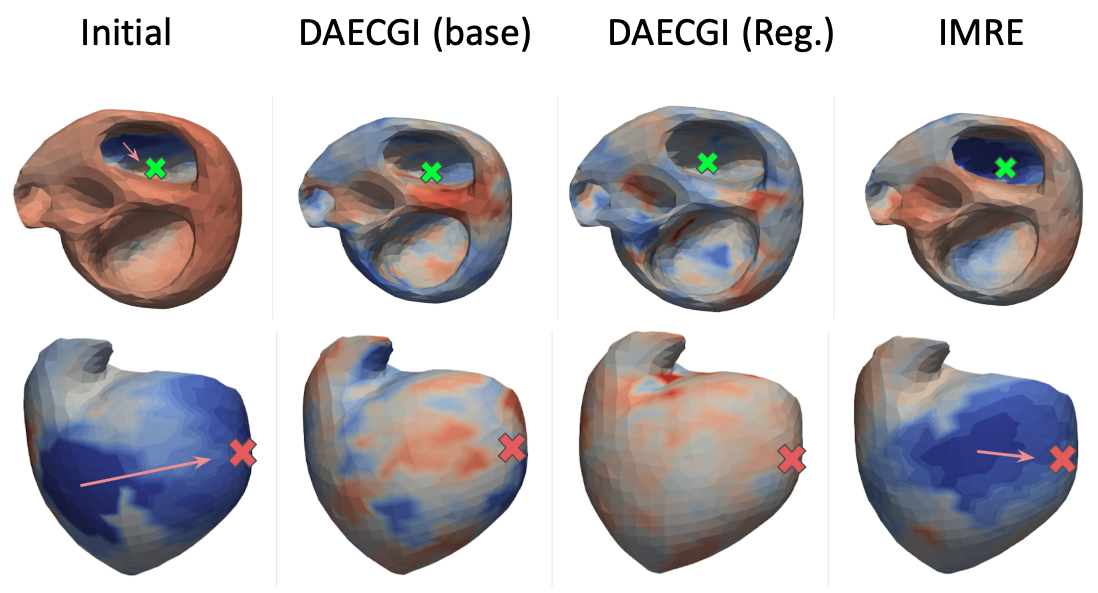}
    \caption{Pacing location localization across two subjects presented in each row. From left to right, Initial ECGi solution, DAECGI with and without regularization, and IMRE final solution. The green and red cross are the ground-truth pacing locations.}
    \label{EDGAR}
\end{figure}

After thorough validation of the generative model, we tested the complete optimization pipeline.

\subsubsection{Synthetic Data Results}: Since we have access to the ground forward operator in the synthetic data, we can easily compare the accuracy of the inverse solutions. Fig. \ref{fig:tcc_scc} compares the spatial and temporal correlation coefficients of inverse solutions obtained from initial forward operator, DAECGI with and without regularization, and IMRE, showing that IMRE outperforms the other methods.
    
\subsubsection{Real Data Results}: As we do not have the ground truth heart surface potential for the real data, we rely on plotting activation points shown in Fig. \ref{EDGAR} to measure the effectiveness of different methods. The euclidean distance between each solution and the ground truth pacing location in Fig. \ref{EDGAR} is; top – initial: 26.3, IMRE: 12.7; bottom – initial: 75.3, IMRE: 34.7. DAECGI failed in producing both results. It is evident from these results that our model shows significant improvement over initial H and DAECGI.

\section{Conclusion}
In this work, motivated by the importance of the physics underlying the forward and inverse imaging problems, we presented Interpretable Modeling and Reduction of
Unknown Errors in Mechanistic Operators (IMRE). IMRE is a interpretable generative-model-based approach to image reconstruction that relies on the forward modeling physics but optimizes the underlying errors to deliver accurate inverse solutions. We showed that by separately modeling errors and forward operators, we can design a more interpretable neural network that leads to more accurate final solutions. This is an initial step towards interpretability in imaging and the current model is still not fully robust in detection of error types. Future work is required to investigate more sophisticated approaches for the interpretability of the latent space of residual errors.  

\section{Acknowledgments}
This work is supported by the National Heart, Lung and Blood Institute (NHLBI) of the National Institutes of Health (NIH) under Award Number R01HL145590.

\bibliographystyle{splncs04}
\bibliography{Ref}

\begin{thebibliography}{10}
\providecommand{\url}[1]{\texttt{#1}}
\providecommand{\urlprefix}{URL }
\providecommand{\doi}[1]{https://doi.org/#1}

\bibitem{ALIEV1996293}
Aliev, R.R., Panfilov, A.V.: A simple two-variable model of cardiac excitation.
  Chaos, Solitons and Fractals  \textbf{7}(3),  293--301 (1996).
  \doi{https://doi.org/10.1016/0960-0779(95)00089-5}

\bibitem{Aras2015}
Aras, K., Good, W., Tate, J., Burton, B., Brooks, D., Coll-Font, J., Doessel,
  O., Schulze, W.H.W., Potyagaylo, D., Wang, L., Dam, P., MacLeod, R.:
  Experimental data and geometric analysis repository-edgar. Journal of
  electrocardiology  \textbf{48} (08 2015).
  \doi{10.1016/j.jelectrocard.2015.08.008}

\bibitem{cartis2018improving}
Cartis, C., Fiala, J., Marteau, B., Roberts, L.: Improving the flexibility and
  robustness of model-based derivative-free optimization solvers (2018)

\bibitem{MRI_ODE}
Chen, E.Z., Chen, T., Sun, S.: Mri image reconstruction via learning
  optimization using neural odes. Lecture Notes in Computer Science p. 83–93
  (2020). \doi{$10.1007/978-3-030-59713-9_9$}

\bibitem{formaggia_quarteroni_veneziani_2006}
Formaggia, L., Quarteroni, A., Veneziani, A.: Complex Systems in Biomedicine.
  Springer-Verlag Italia, Milano (2006)

\bibitem{Ghimire}
Ghimire, S., Dhamala, J., Gyawali, P., Sapp, J., Horacek, B., Wang, L.:
  Generative modeling and inverse imaging of cardiac transmembrane potential
  (05 2019)

\bibitem{Gulrajani}
Gulrajani, R.: The forward and inverse problems of electrocardiography. IEEE
  Engineering in Medicine and Biology Magazine  \textbf{17}(5),  84--101
  (1998). \doi{10.1109/51.715491}

\bibitem{Horek1997TheIP}
Hor{\'a}ek, B.M., Clements, J.C.: The inverse problem of electrocardiography: a
  solution in terms of single- and double-layer sources of the epicardial
  surface. Mathematical biosciences  \textbf{144 2},  119--54 (1997)

\bibitem{DeepPET}
Häggström, I., Schmidtlein, C., Campanella, G., Fuchs, T.: Deeppet: A deep
  encoder-decoder network for directly solving the pet image reconstruction
  inverse problem. Medical Image Analysis  \textbf{54},  253--262 (03 2019)

\bibitem{SCI:SCIRun}
Institute, S.:  (2016), sCIRun: A Scientific Computing Problem Solving
  Environment, Scientific Computing and Imaging Institute (SCI), Download from:
  http://www.scirun.org

\bibitem{kohonen1990self}
Kohonen, T.: The self-organizing map. Proceedings of the IEEE  \textbf{78}(9),
  1464--1480 (1990)

\bibitem{Proximal}
Lai, K.W., Aggarwal, M., van Zijl, P., Li, X., Sulam, J.: Learned proximal
  networks for quantitative susceptibility mapping. In: Medical Image Computing
  and Computer Assisted Intervention – MICCAI 2020: 23rd International
  Conference, Lima, Peru, October 4–8, 2020, Proceedings, Part II. p.
  125–135. Springer-Verlag, Berlin, Heidelberg (2020).
  \doi{$10.1007/978-3-030-59713-9_13$}

\bibitem{8253590}
Lucas, A., Iliadis, M., Molina, R., Katsaggelos, A.K.: Using deep neural
  networks for inverse problems in imaging: Beyond analytical methods. IEEE
  Signal Processing Magazine  \textbf{35}(1),  20--36 (2018).
  \doi{10.1109/MSP.2017.2760358}

\bibitem{natterer2001mathematical}
Natterer, F., W{\"u}bbeling, F.: Mathematical methods in image reconstruction.
  SIAM (2001)

\bibitem{plonsey_fleming_1989}
Plonsey, R., Fleming, D.G.: Bioelectric phenomena. McGraw-Hill (1989)

\bibitem{potyagaylo2019ecg}
Potyagaylo, D., Chmelevsky, M., Van~Dam, P., Budanova, M., Zubarev, S.,
  Treshkur, T., Lebedev, D.: Ecg adapted fastest route algorithm to localize
  the ectopic excitation origin in crt patients. Frontiers in Physiology
  \textbf{10}, ~183 (2019)

\bibitem{ramanarayanan_murugesan_ram_sivaprakasam_2020}
Ramanarayanan, S., Murugesan, B., Ram, K., Sivaprakasam, M.: Dc-wcnn: A deep
  cascade of wavelet based convolutional neural networks for mr image
  reconstruction. 2020 IEEE 17th International Symposium on Biomedical Imaging
  (ISBI)  (2020). \doi{10.1109/isbi45749.2020.9098491}

\bibitem{UNET}
Ronneberger, O., Fischer, P., Brox, T.: U-net: Convolutional networks for
  biomedical image segmentation. In: Navab, N., Hornegger, J., Wells, W.M.,
  Frangi, A.F. (eds.) Medical Image Computing and Computer-Assisted
  Intervention -- MICCAI 2015. pp. 234--241. Springer International Publishing,
  Cham (2015)

\bibitem{476126}
Throne, R., Olson, L.: The effects of errors in assumed conductivities and
  geometry on numerical solutions to the inverse problem of
  electrocardiography. IEEE Transactions on Biomedical Engineering
  \textbf{42}(12),  1192--1200 (1995). \doi{10.1109/10.476126}

\bibitem{MTBDAECGI2021}
Toloubidokhti, M., Gyawali, P.K., Gharbia, O.A., Jiang, X., Font, J.C.,
  Bergquist, J.A., Zenger, B., Good, W.W., Brooks, D.H., MacLeod, R.S., Wang,
  L.: Deep adaptive electrocardiographic imaging with generative forward model
  for error reduction. In: Ennis, D.B., Perotti, L.E., Wang, V.Y. (eds.)
  Functional Imaging and Modeling of the Heart. pp. 471--481. Springer
  International Publishing, Cham (2021)

\bibitem{vettigliminisom}
Vettigli, G.: Minisom: minimalistic and numpy-based implementation of the self
  organizing map (2018), \url{https://github.com/JustGlowing/minisom/}

\end{thebibliography}

\newpage
\section{Supplementary materials}

\subsection{Convergence plots for inverse estimation}

\begin{figure}
    \centering
    \includegraphics[width=0.7\textwidth, height=5.2cm]{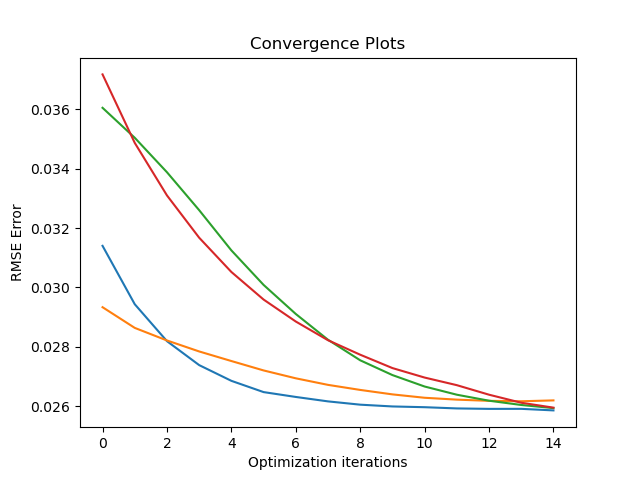}
    \caption{Reducing RMSE error between predicted inverse solution and ground truth on simulation data over optimization iterations. Each curve is one example.}
    \label{fig:convergence_plots}
\end{figure}
\subsection{Visualization of heart potential}
\begin{figure}
    \centering
    \includegraphics[width=\textwidth,scale=1]{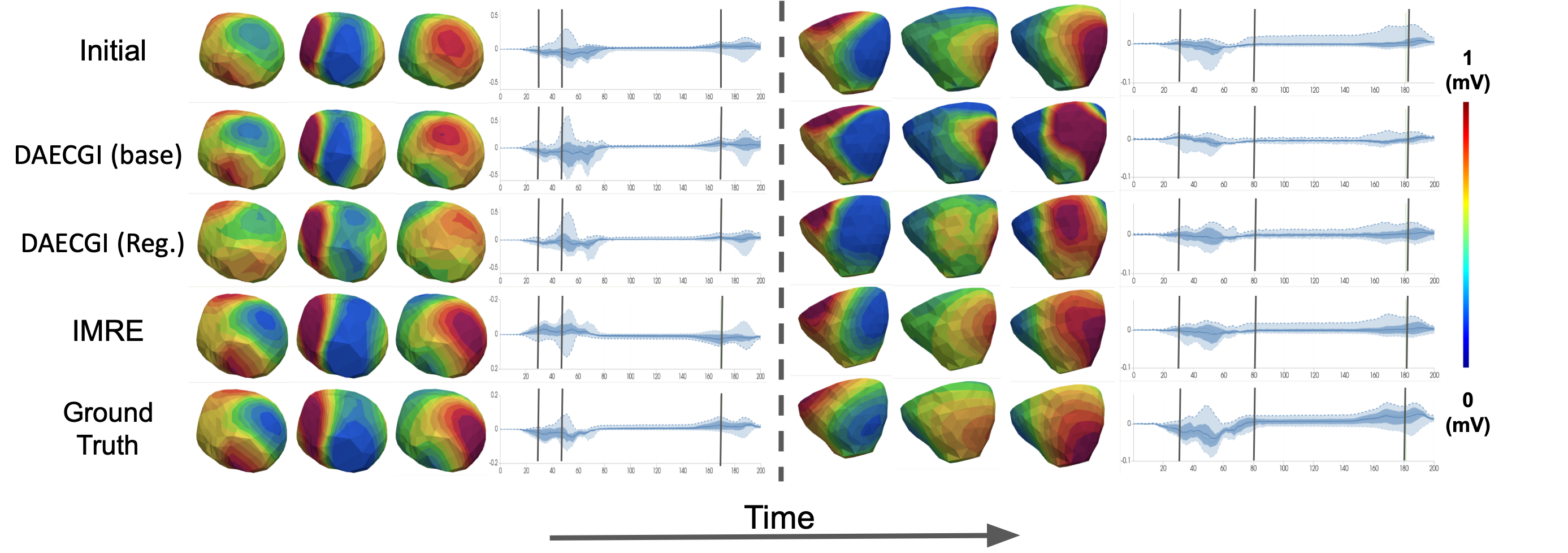}
    \caption{Examples of heart electrical potentials generated from each model after the optimization stage. The dashed bar separates the two examples. In each example the vertical lines showcase the time stamps of the displayed potential distribution.}
    \label{fig:convergence_plots}
\end{figure}

\end{document}